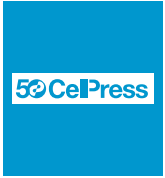
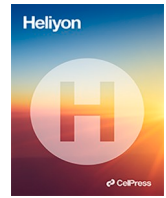

Research article

# Estimated electric conductivities of thermal plasma for air-fuel combustion and oxy-fuel combustion with potassium or cesium seeding

Osama A. Marzouk

*College of Engineering, University of Buraimi, Al Buraimi, Sultanate of Oman*

A B S T R A C T

A complete model for estimating the electric conductivity of combustion product gases, with added cesium (Cs) or potassium (K) vapor for ionization, is presented. Neutral carrier gases serve as the bulk fluid that carries the seed material, as well as the electrons generated by the partial thermal (equilibrium) ionization of the seed alkali metal. The model accounts for electron-neutral scattering, as well as electron-ion and electron-electron scattering. The model is tested through comparison with published data. The model is aimed at being utilized for the plasma within magnetohydrodynamic (MHD) channels, where direct power extraction from passing electrically conducting plasma gas enables electric power generation.

The thermal ionization model is then used to estimate the electric conductivity of seeded combustion gases under complete combustion of three selected fuels, namely: hydrogen ($H_2$), methane ($CH_4$), and carbon (C). For each of these three fuels, two options for the oxidizer were applied, namely: air (21 % molecular oxygen, 79 % molecular nitrogen by mole), and pure oxygen (oxy-combustion). Two types of seeds (with 1 % mole fraction, based on the composition before ionization) were also applied for each of the six combinations of (fuel-oxidizer), leading to a total of 12 different MHD plasma cases. For each of these cases, the electric conductivity was computed for a range of temperatures from 2000 K to 3000 K.

The smallest estimated electric conductivity was 0.35 S/m for oxy-hydrogen combustion at 2000 K, with potassium seeding. The largest estimated electric conductivity was 180.30 S/m for oxy-carbon combustion at 3000 K, with cesium seeding. At 2000 K, replacing potassium with cesium causes a gain in the electric conductivity by a multiplicative gain factor of about 3.6 regardless of the fuel and oxidizer. This gain factor declines to between 1.77 and 2.07 at 3000 K.

Based on the findings of this research study, the four analyzed factors to increase the electric conductivity of MHD plasma can be listed by their significance (descending order) as (1) type of additive seed type (cesium is better than potassium), (2) temperature (the higher the better), (3) carbon-to-hydrogen ratio of the fuel (the higher the better), and finally (4) the oxidizer type (air is generally better than pure oxygen).

The relative size of the two electric conductivity components (due to neutrals scattering and Coulomb scattering) at various plasma conditions are discussed, and a threshold of $10^{-5}$ (0.001 %) electrons mole fraction is suggested to safely neglect Coulomb scattering.

**Nomenclature**

| | |
|---|---|
| $\alpha$ | Degree of ionization, the percentage of the alkali metal atoms that got ionized [%] |
| $\varepsilon_0$ | Electric permittivity in a vacuum, $8.8541878128 \times 10^{-12}$ [F/m] [1] |
| $\epsilon_i$ | Ionization energy or ionization potential (for first-level ionization), the energy needed to remove an electron from the outermost electronic orbit [eV] |
| $\Lambda$ | Computed quantity (cutoff parameter), which depends on the temperature and electron density. Its natural logarithm, $\ln(\Lambda)$, is called the Coulomb logarithm |
| $\mu$ | Estimated electron mobility in plasma, taking into account both neutrals scattering and Coulomb scattering [m$^2$/s.V] |
| $\nu$ | Overall mean collision cyclic frequency due to the combined effect of neutrals scattering and Coulomb scattering, $\nu = \nu_0 + \nu_1$ [Hz, 1/s] |
| $\nu_0$ | Mean collision cyclic frequency due to neutrals scattering [Hz, 1/s] |
| $\nu_1$ | Mean collision cyclic frequency due to Coulomb scattering [Hz, 1/s] |
| $\theta$ | Thermal energy, transformed absolute temperature of plasma by expressing it as electronvolts of energy, $\theta = \widehat{k}\,T$ [eV] |
| $\sigma$ | Overall electric conductivity of plasma due to the combined effect of neutrals scattering and Coulomb scattering [S/m] |
| $\sigma_0$ | Electric conductivity of plasma due to only neutrals scattering [S/m] |
| $\sigma_1$ | Electric conductivity of plasma due to only Coulomb scattering [S/m] |
| $\tau$ | Mean time between electron collisions, taking into account both neutrals scattering and Coulomb scattering [s] |
| $\xi$ | Intermediate temperature-dependent variable used in the Maxwell-Boltzmann distribution, $\xi = m/(2k\,T)$ [kg/J] |
| $\zeta$ | Overall electric resistivity of plasma due to the combined effect of neutrals scattering and Coulomb scattering, $\zeta = \zeta_0 + \zeta_1$ [m/S] |
| $\zeta_0$ | Electric resistivity of plasma due to neutrals scattering, $\zeta_0 = 1/\sigma_0$ [m/S] |
| $\zeta_1$ | Electric resistivity of plasma due to Coulomb scattering, $\zeta_1 = 1/\sigma_1$ [m/S] |
| $C$ | Electron speed in Maxwell-Boltzmann distribution [m/s] |
| $C_{\varepsilon 0}$ | Coulomb constant, a derived constant that appears in the formula of $C_\Lambda$ and $C_{K\nu 1}$, $C_\varepsilon = 1/(4\,\pi\,\varepsilon_0) = 8.987551792261 \times 10^9$ [m/F] |
| $C_\Lambda$ | Derived constant that appears in the formula of $\Lambda$, $1.5487761 \times 10^{13}$ [1/(V$^{1.5}$.m$^{1.5}$)] |
| $C_{K\nu 1}$ | Derived constant that appears in the formula of $\nu_1$, $7.72696841 \times 10^{-12}$ [V$^2$.m$^2$.C$^{0.5}$/kg$^{0.5}$] |
| $C_{Saha}$ | Derived constant that appears in the formula of $K_{Saha}$, $2.4146830 \times 10^{21}$ [1/(K$^{1.5}$.m$^3$)] |
| $C_v$ | Derived constant that appears in the formula of the mean electron speed, 6212.511428620 [m/(s.K$^{0.5}$))] |
| $e$ | Electron charge magnitude, also called elementary charge, $1.602176634 \times 10^{-19}$ [C] [2] |
| eV | Electronvolt or electron volt, $1.602176634 \times 10^{-19}$ [J] [3] |
| $h$ | Planck constant, $6.62607015 \times 10^{-34}$ [J/K] [4] |
| $K$ | Electron kinetic energy corresponding to a given speed for the Maxwell-Boltzmann distribution, when expressed in electronvolts, $K = \mathbf{0.5\,m\,C^2}/e$ [eV] |
| $K_{Saha}$ | An intermediate quantity (the ionization equilibrium constant) for computing $n_e$, the equilibrium constant for ionization of alkali metal atoms. For a given alkali metal seed, it depends only on the absolute temperature of the plasma. [1/m$^3$] |
| $k$ | Boltzmann constant, $1.380649 \times 10^{-23}$ [J/Hz, J.s] [5] |
| $\widehat{k}$ | Transformed Boltzmann constant, $\widehat{k} = k/e = 8.617333262 \times 10^{-5}$ [eV/K] [6] |
| $m$ | Electron mass, $9.1093837015 \times 10^{-31}$ [kg] [7] |
| $n_e$ | Number density of free electrons as detached particles in a plasma, after ionization. Because free electrons are present only after ionization (no free electrons before ionization), the use of a single prime symbol and a double prime symbol to distinguish between these two states is redundant, and thus is not adopted. [1/m$^3$] |
| $n_i$ | Number density of ions in the plasma. Here: $n_i = n_e$ because the seeded alkali metal atoms are singly ionized (each electron is liberated from a single seed atom). Because ions are present only after ionization (no ions before ionization), the use of a single prime symbol and a double prime symbol to distinguish between these two states is redundant, and thus is not adopted. [1/m$^3$] |
| $n'_s$ | Number density of seed atoms before ionization [1/m$^3$] |
| $n''_s$ | Number density of seed atoms after ionization, $n''_s = n'_s - n_e$ [1/m$^3$] |
| $n'_{tot}$ | Number density of all particles existing in the plasma before ionization (carrier molecules or atoms and seed atoms) [1/m$^3$] |
| $n''_{tot}$ | Number density of all particles existing in the plasma after ionization (carrier molecules or atoms, seed atoms, seed ions, and free electrons), $n''_{tot} = n'_{tot} + n_e$ [1/m$^3$] |
| $p_e$ | Partial static pressure of free electrons after ionization. Because free electrons are present only after ionization (no free electrons before ionization), the use of a single prime symbol and a double prime symbol to distinguish between these two states is redundant, and thus is not adopted. [Pa] |
| $p'_j$ | Partial static pressure of the jth neutral species (carriers or atomic seed) before ionization [Pa] |
| $p'_s$ | Partial static pressure of the seeded alkali metal vapor before ionization, the "seeding pressure", $p'_s = X'_s \, p'_{tot}$ [Pa] |
| $p'_{tot}$ | Total static pressure of the plasma mixture before ionization, $p'_{tot} = \sum_j p'_j$ [Pa] |
| $p''_{tot}$ | Total static pressure of the plasma mixture after ionization, $p''_{tot} = p'_{tot} + p_e$ [Pa] |
| $Q$ | Mean electron-neutral collision cross-section, for a particular neutral species [Å$^2$] |
| $T$ | Absolute temperature of the plasma, same for electrons and heavy particles [K] |
| $u$ | Mean kinetic energy of the electron, expressed in electronvolts [eV] |
| $v$ | Mean electron speed, according to the Maxwell-Boltzmann distribution [m/s] |
| $X_e$ | Mole fraction of ions after ionization. Because free electrons are present only after ionization (no free electrons before ionization), the use of a single prime symbol and a double prime symbol to distinguish between these two states is redundant and is not adopted. [−] |
| $X'_s$ | Mole fraction of seeded alkali metal vapor before ionization [−] |

## 1. Introduction

### 1.1. Background

In order to achieve global carbon neutrality (net-zero emissions of carbon dioxide) by 2050, and thus limit the rise in the global mean surface temperature (compared to the pre-industrial level) to 1.5 °C by 2100, total electricity generation in 2050 should grow to about three times its level in 2020, with the electrification percentage in the total final energy consumption (TFEC) reaching about 50 % in 2050 (compared to 22 % in 2020), and the share of renewable energy sources should dominate total electricity generation in 2050, with a share of about 90 % in 2050 (compared to 28 % in 2020) [8–13]. While the increasing demand for electricity can be fulfilled by conventional fossil-fuel power plants, nuclear power plants, and mature renewable energy power plants (such as solar photovoltaic panels and wind turbines) [14–24], the current study is concerned with a special electricity generation method, which is magnetohydrodynamic (MHD) channels (or plasma generators).

Direct power extraction (DPE) from the flowing high-speed plasma gas within a magnetohydrodynamic (MHD) channel [25–31] is a legacy concept that was explored for about three decades after the energy crisis in 1973–1974 [32–36], and even before that in the 1960s. These endeavors proved that MHD power generation is viable, but also is economically non-competitive and technologically challenging. However, with advancements in related technologies, such as superconducting electromagnets [37–41], and the alignment of MHD power generation with the elevated-temperature oxygen-fired power plants coupled with carbon capture for the $CO_2$-enriched combustion products [42–54], this concept was revisited recently with a potential for realization or at least conducting revised feasibility studies [55,56].

Plasma, as an electrically conducting ionized gas, is overall neutral since the free electrons within it originate from local atoms or molecules. There are different ways to ionize plasma. The one considered here is thermal ionization, where electrons are liberated from heavy neutral particles (molecules or atoms) at high temperatures due to sufficiently-energetic collisions between two heavy particles, or between a heavy particle and a liberated electron. The kinetic energy of the incident particle or electron must be high enough such that the energy transferred in the collision is equal to or greater than the ionization energy, in order to separate an electron from a heavy particle. Thus, thermal ionization is collisional ionization. Electrons are much more efficient than heavy particles as ionizing agents, and thus electron collisional ionization dominates the thermal ionization process [57].

For an ideal gas in thermal equilibrium (having a single uniform temperature), the speed (the velocity magnitude) of particles can be described by the Maxwell-Boltzmann probability distribution function [58–60]. This distribution (expressed as the probability per unit speed) for the thermal speed of particles depends on the temperature ($T$), where at higher temperatures, electrons tend to have higher speeds. The Maxwell-Boltzmann distribution for speeds of free plasma electrons (liberated electrons, not those orbiting within an atom) is illustrated in Fig. 1 at four temperatures. Its mathematical form is shown in Equation (1).

$$f(C;T) = \frac{4}{\sqrt{\pi}}(\xi)^{1.5} \, C^2 \, e^{-\xi \, C^2} \; ; \; \xi[\text{kg}/\text{J}] \equiv \frac{m}{2k\,T} \qquad (1)$$

where ($C$) is the electron speed as the variable of the distribution, and ($T$) is the absolute temperature of the plasma gas. For this distribution, the area under the curve within a certain speed range gives the estimated fraction of electrons possessing speeds within that speed range. The Maxwell-Boltzmann distribution can be viewed as a function of probable electron kinetic energy expressed in electronvolts ($K[\text{eV}]$), which is related to the electron speed of the distribution ($C$) as shown in Equation (2).

$$K[\text{eV}] = \frac{1}{2}\frac{m\,C^2}{e} = 2.8428150518 \times 10^{-12} \, C^2 \, [\text{m}^2/\text{s}^2] \qquad (2)$$

Thus, the corresponding range of electron kinetic energy ($K[\text{eV}]$) is added in the figure.

Thermal ionization by heating results in a condition of thermal equilibrium, with both the electrons and the heavy particles having the same temperature [61]. This is unlike ionization established by electric glow discharge in fluorescent lighting tubes [62,63], where electrons have a temperature that is higher than the temperature of ions and other heavy particles. In MHD plasma based on combustion product gases (flue gases), electrons and heavy particles are in thermal equilibrium due to high collision frequencies and energy exchange per collision [64], leading to a "thermalization" process by which the temperature becomes uniformly distributed [65].

Thermal equilibrium plasma (single temperature plasma) achieved by heating is suitable for MHD channels, where the high temperature of combustion product gases is readily available to partly ionize the flue gases such that a sufficient level of electric conductivity is attained. However, examining the ionization energy of common combustion gases (such as carbon dioxide and water vapor) in comparison to alkali metals reveals that ionizing combustion gases is not possible even at elevated temperatures of oxy-fuel combustions near 3000 K. With oxy-fuel combustion (or oxygen combustion), the oxidizer is pure oxygen, rather than oxygen diluted with nitrogen and argon as in the case of conventional air-combustion. This absence of non-reacting gases (that would absorb a part of the released combustion heat) causes the combustion temperature to rise [66–73]. Even at 6000 K, no appreciable ionization is expected due to simple thermal ionization of combustion gases [74] if an atmospheric pressure is maintained. Reducing the pressure to near vacuum levels helps in increasing the ionization and thus the electrical conductivity of plasma [75]. However, such a very low pressure is not adequate for combustion plasma, and also the gain in ionization is not high enough to allow suitable numbers of electrons at the typical range of combustion temperatures, and even at a higher temperature of 4000 K [76]. In order to achieve useful electric conductivity by thermal equilibrium ionization in combustion plasma, a small amount (such as 1 % by mass or even less) of alkali metal seeding material is needed. Cesium (Cs) and potassium (K) alkali metals are particularly recommended seeding elements because they have low ionization energies [77]. Additionally, they have relatively low boiling temperatures, making them vapors at MHD operating temperatures, thus they mix smoothly with other plasma gases without having two-phase gas-solid or gas-liquid flows. Cesium has the lowest ionization energy among all chemical elements in the periodic table. Francium has the second smallest ionization energy, but it is extremely rare in nature, being the second-rarest natural element after the radioactive element astatine (At) [78], and it is also radioactive and very unstable [79–81]. Rubidium (Rb) has the third smallest ionization energy, and potassium has the fourth smallest ionization energy. Potassium has the advantage of being less expensive and more abundant than cesium, francium, and rubidium. Potassium is the seventh most abundant element in the continental crust of the Earth, making up about 2.5 % by mass of the Earth's crust [82,83]. Table 1 lists the ionization energy and the boiling temperature of these four alkali metals, as well as of the remaining two alkali metals (sodium, Na; and lithium, Li) in the periodic table. In addition, eight other elements are listed for comparison [84,85]. The table includes helium (He) at its end, which is the element having the largest ionization energy in the periodic table, thus it is the least ionizable element.

Due to the big difference between the energy of the first ionization (level 1) and the energy of the second ionization (level 2) of alkali metals and of atoms in general [86], only the first ionization is considered here. For example, the first ionization of potassium (level 1, from an ion to a singly-ionized ion with an electric charge of $e$) has a potential of 4.34 eV, which jumps to 31.63 eV at the second ionization (level 2, from a singly-ionized ion to a doubly-ionized ion with a charge of $2e$) [87]. Similarly, the first ionization energy of cesium is 3.89 eV, which is much smaller than its second ionization energy of 23.16 eV [88]. Thus, the seeded alkali metal atoms are singly-ionized (one atom can lose only one electron at maximum).

There are two extreme cases of plasma gases, which are the very-weakly-ionized plasma, and the fully-ionized plasma. This division is based on the relative contribution of the two scattering modes of free liberated electrons acting as the effective charge carriers in the electrically-conducting plasma gas, since ions are relatively immobile compared to electrons [89]. In the very-weakly-ionized plasma, the ionization level is very small such that the number density of ions (and thus the number density of electrons) is less than about $10^{-4}$ (0.01 %) of the number density of neutrals, so the Coulomb scattering (interaction between electrons and ions or other electrons) may be neglected and only the neutrals scattering is important. It should be noted here that ions have about three orders of magnitude

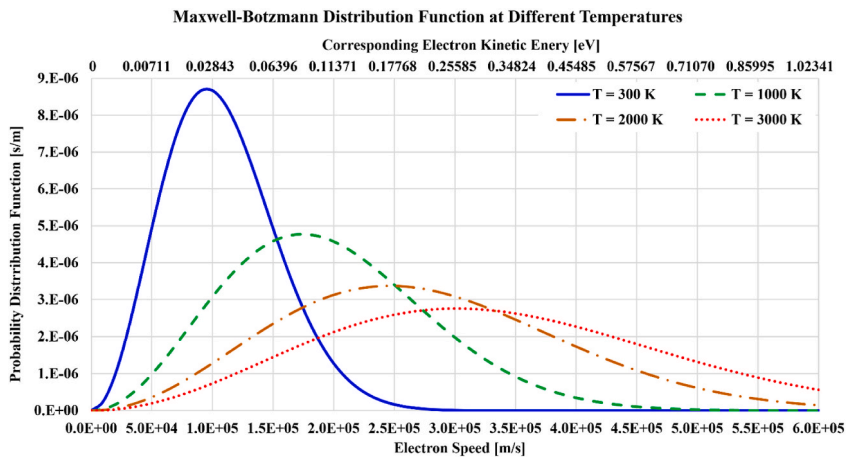

**Fig. 1.** Illustration of Maxwell-Boltzmann speed probability distribution of electrons at four absolute temperatures: 300 K, 1000 K, 2000 K, and 3000 K.

**Table 1**

Ionization energy and boiling temperature (boiling point) of the 6 alkali metals, compared with other 8 elements. The elements are ordered ascendingly by the ionization energy.

| Species Name (and Symbol) | Atomic Number | Electron Configuration | Ionization Energy [eV] | Boiling Temperature [K] |
| --- | --- | --- | --- | --- |
| **Cesium (Cs)** | 55 | [Xe]$6s^1$ | 3.89390572743 | 944 |
| **Francium (Fr)** | 87 | [Rn]$7s^1$ | 4.0727411 | 923 |
| **Rubidium (Rb)** | 37 | [Kr]$5s^1$ | 4.1771281 | 961 |
| **Potassium (K)** | 19 | [Ar]$4s^1$ | 4.34066373 | 1032 |
| **Sodium (Na)** | 11 | [Ne]$3s^1$ | 5.13907696 | 1156 |
| **Lithium (Li)** | 3 | $1s^2 2s^1$ | 5.391714996 | 1615 |
| **Aluminum (Al)** | 13 | [Ne]$3s^2 3p^1$ | 5.985769 | 2792 |
| **Calcium (Ca)** | 20 | [Ar]$4s^2$ | 6.11315547 | 1757 |
| **Magnesium (Mg)** | 12 | [Ne]$3s^2$ | 7.646236 | 1363 |
| **Copper (Cu)** | 29 | [Ar]$3d^{10} 4s^1$ | 7.726380 | 2833 |
| **Beryllium (Be)** | 4 | $1s^2 2s^2$ | 9.322699 | 2741 |
| **Argon (Ar)** | 18 | [Ne]$3s^2 3p^6$ | 15.7596119 | 87.302 |
| **Neon (Ne)** | 10 | $1s^2 2s^2 2p^6$ | 21.564541 | 27.104 |
| **Helium (He)** | 2 | $1s^2$ | 24.587389011 | 4.222 |

stronger interaction with electrons than neutrals, which justifies why a very low ion concentration is necessary to ignore their hindrance to electron mobility [90]. The other extreme case of plasma in terms of ionization level is the fully-ionized plasma, which contains no neutrals, but consists of ions and electrons only. In that case, the neutrals scattering vanishes and only Coulomb scattering exists. Neither extreme case of plasma is suitable for direct power extraction through MHD channels. In the very-weakly-ionized plasma, the number of available free electrons is deficient, causing the electric conductivity to be too small for a useful electric output. In the fully-ionized plasma, impractically-high temperatures are required to ionize all neutrals. Instead, a partially-ionized (or weakly-ionized) plasma is sought for direct power extraction using MHD channels, with a small ionization level and with Coulomb scattering also present. Weak ionization is preferred because if the ionization level increases to a high level, the Coulomb scattering grows to a level that does not cause significant improvement in the plasma electric conductivity. Thus, the number density of electrons is not recommended to exceed 0.001 (0.1 %) of the number density of neutrals in the plasma [91]. With such weak ionization, the use of ideal gas law is valid, and is not violated by the small fraction of electrons added to the neutral particles and the small fraction of neutrals that turn into ions [92,93].

To account for the influence of both modes of scattering (neutrals scattering and Coulomb scattering) on the plasma electric conductivity, the resistivity-addition concept can be used [94–96], such that

$$\zeta \equiv \zeta_0 + \zeta_1 \tag{3}$$

or

$$\frac{1}{\sigma} \equiv \frac{1}{\sigma_0} + \frac{1}{\sigma_1} \tag{4}$$

Mathematically, Equation (4) indicates that the overall electric conductivity is half of the harmonic mean [97,98] of the two constituent electric conductivities, or

$$\sigma = \frac{1}{1/\sigma_0 + 1/\sigma_1} = \frac{\sigma_0 \, \sigma_1}{\sigma_0 + \sigma_1} = \frac{1}{\zeta_0 + \zeta_1} = \frac{1}{\zeta} \tag{5}$$

Equation (5) also indicates that the overall electric conductivity ($\sigma$) is less than the smaller of ($\sigma_0$) and ($\sigma_1$).

The two constituent electric resistivities take the form

$$\zeta_0 = \frac{m \, \nu_0}{n_e \, e^2} \; ; \; \zeta_1 = \frac{m \, \nu_1}{n_e \, e^2} \tag{6}$$

As the reciprocal of electric resistivities, the two constituent electric conductivities take the form shown in Equation (7).

$$\sigma_0 = \frac{n_e \, e^2}{m \, \nu_0} \; ; \; \sigma_1 = \frac{n_e \, e^2}{m \, \nu_1} \tag{7}$$

Using Equation (6) in Equation (3) gives Equation (8), which is

$$\zeta = \frac{m \, \nu_0}{n_e \, e^2} + \frac{m \, \nu_1}{n_e \, e^2} \tag{8}$$

Because ($m$) and ($e$) are physical constants, and ($n_e$) is a common plasma property regardless of the mode of scattering, the overall electric resistivity or the overall electric conductivity of the plasma is effectively a function of the sum of the two mean collision frequencies, as described in Equations 9–11 [99].

$$\zeta = \frac{m}{n_e \, e^2} \, (\nu_0 + \nu_1) \tag{9}$$

$$\zeta = \frac{m}{n_e \, e^2} \nu = \frac{m}{n_e \, e^2 \, \tau} \tag{10}$$

$$\sigma = \frac{n_e \, e^2}{m \, \nu} = \frac{n_e \, e^2 \, \tau}{m} \tag{11}$$

Therefore, the core process of computing the electric conductivity of partially-ionized plasma with an arbitrary level of ionization is estimating the two mean collision frequencies describing the neutral-electron collisions and the Coulomb scattering of electrons. The computations involve nonlinear and complex equations, making it difficult to recognize the dependence of the overall electric conductivity on the (1) plasma temperature, (2) the seed material type (as reflected particularly in its ionization energy value and its electron-neutral collision cross-section), and the (3) carrier neutral gases in the plasma (as reflected particularly in their electron-neutral collision cross-sections). Therefore, computerized simulations and visualized results effectively reveal the influence of these three plasma parameters. This is the main objective of the present study.

*1.2. Goal and structure of the study*

The present study aims at applying a physical-mathematical model for estimating the electric conductivity of a partially-ionized plasma gas pertaining to MHD direct power extraction. The model is largely based on formulas in the literature, with a simplifying change and with the use of a novel nonlinear regression function for the argon gas (allowing convenient computation of its electron-neutral collision cross-section at a given electron speed and a given temperature). The source of electrons in the modeled plasma is the thermal ionization of either cesium or potassium vapor, added with a fixed mole fraction of 0.01 (1 %) of the pre-ionization gas mixture. The pressure is atmospheric (absolute pressure of 1 standard atmosphere) before ionization and after the seed is added. The ability of the model to estimate the electron number density and the electric conductivity was tested through comparisons with published data. During this step, an eighth-order polynomial was suggested for the electron-neutral collision cross-section of argon, as a function of the electron kinetic energy. The model is used to inspect the gain expected when switching from potassium to cesium in two electric characteristics of the plasma that are independent of the chemical composition of the carrier (non-seed) gases. These electric characteristics are the electron density, and the degree of ionization. The range of temperatures considered is from 2000 K to 3000 K. Then, the variations of the plasma electric conductivity with the temperature, and the gain expected due to changing the seed type from potassium to cesium, are provided for the same temperature range. Six different stoichiometric (no excess fuel, no excess oxidizer) combustion scenarios were numerically simulated, which correspond to three fuel options and two oxidizer options. The three fuels examined are molecular hydrogen, methane, and carbon. The two oxidizers examined are oxygen and air (oxygen-nitrogen mixture). This allows identifying the effect of the chemical composition of the flue gas to be seeded, ranging from pure water vapor to pure carbon dioxide. Finally, the relative size of the neutrals-scattering-based electric conductivity component and the Coulomb scattering-based electric conductivity component is discussed at the lower, middle, and upper temperatures of the range of interest (thus at 2000 K, 2500 K, and 3000 K), for each of the six combustion scenarios.

While the subject addressed by the current study is not new, the current study complements previous similar studies that have already examined the influence of other plasma parameters on the electric conductivity, such as the seed amount, the total static pressure, and the fuel-air ratio [100,101]. The presented study provides new results and is focused on MHD applications. It provides qualitative and quantitative information about electric conductivities that can guide the design of a magnetohydrodynamic (MHD) channel, such as the selection of the seed type by balancing its performance gain against its added cost and handling concerns, and the selection of desired temperature or chemical composition for the flue gas. The study is not limited to a single particular fuel-oxidizer condition, but covers multiple generic scenarios, which makes its findings broad yet informative.

Aside from the results given here for elucidating how the electric conductivity of seeded plasma varies in response to different factors, the detailed presentation of a procedure to numerically compute the electric conductivity at a point, given the local temperature, total pressure, and chemical composition can be useful for computational fluid dynamics (CFD) models that require such an electric conductivity submodel (a submodel is one part of the whole modeling process), that is combined with other submodels for enabling a complete numerical simulation of the plasma gas and thus estimating the amount of electric power output from a magnetohydrodynamic (MHD) generator system [102–114].

**2. Research method**

In the current section, the model utilized for estimating the electric conductivity of alkali-metal-seeded, thermal equilibrium (single temperature), and partially-ionized plasma gas is described. The model is largely based on the Frost model [115,116]. In this computational model, a procedure with mathematical formulas is proposed to find the electric conductivity of a gas mixture (which forms partially-ionized high-temperature plasma) seeded with a typical plasma seeding gas for magnetohydrodynamic electric power generation. Through these formulas, the electron mobility in the gas mixture is obtained, taking into account the variation of the scattering cross-section with energy. The model utilizes an empirical rule for combining the effects of electron scattering by neutrals and electron scattering by ions. The Frost model was used here as a basis for the analysis due to its comprehensiveness, rigorous details presented for its development, and the numerous literature resources incorporated in it. The Frost model does not neglect electron scattering by ions, thus it is not limited to very-weakly-ionized plasma, where the effect of ions can be neglected. The Frost model is not limited to a specific gaseous species. Instead, it can accommodate any gaseous species as a constituent in the plasma gas mixture, provided that its analytic expression describes how its neutrals-only scattering behavior changes with energy, as a polynomial function. Furthermore, the Frost model is carefully devised such that it gives the correct electric conductivity in the limiting case of completely-ionized plasma. The Frost model was used in different studies about plasma [117–119]. However, due to its relative complexity, the Frost model was modified (simplified) here for more-practical utilization as a standalone submodel within a larger numerical solver for computational fluid dynamics (CFD). The simplification presented here eliminates the dependence on the energy variable, and substitutes this with dependence on the local temperature.

For some formulas, the units (SI units are used here) have been specified between square brackets. This helps in clarifying the calculation process and makes it easily reproducible by others. Similarly, the numerical values of some derived constants encountered during the computation process are provided.

### 2.1. Assumptions

The following assumptions are incorporated in the performed modeling applied in the current study for estimating the electric conductivity of plasma.

- The ionizable material is only the seeded alkali metal, which is singly ionized. Ionization of other carriers (non-seed gaseous species in the combustion product gases) is neglected.
- The seeded alkali metal is introduced to the combustion product gases as an elemental vapor.
- Ionization reflects an equilibrium process (plasma has a single temperature, and transient changes till equilibrium occurs are not captured).
- Energy(or speed)-dependent computations (such as computing the electron-neutral collision cross-sections) are performed at a single temperature-dependent electron kinetic energy value of $u[\text{eV}] = 1.5 \, \theta[\text{eV}] = 1.5 \, \hat{k}_B[\text{eV}/\text{K}] \, T[\text{K}] = 0.00012925999893 \, T[\text{K}]$, which corresponds to the mean electron kinetic energy according to the Maxwell-Boltzmann distribution [120].
- The plasma behaves as an ideal gas.

### 2.2. Calculating the number density of electrons

The number density of electrons from the seed alkali atoms is determined using the Saha equation [121,122]. The Saha equation for the ionization of a gas in thermal equilibrium is a mathematical expression that describes the relation between the ionization state (ions density) of that gas and the temperature and pressure [123–126]. For alkali metals, the equation relates the number density of seed atoms that got ionized (thus, the number density of electrons) and the number density of seed atoms that remained neutral after the equilibrium thermal ionization is completed. This alkali metal version of the Saha equation has the form

$$K_{\text{Saha}} \left[\frac{1}{m^3}\right] \equiv \frac{n_e \, n_i}{n_s''} = C_{\text{Saha}} \, T^{1.5} \, e^{-\frac{\epsilon_i}{\theta}} \tag{12}$$

where ($\epsilon_i$) is the ionization energy (in eV) of the alkali metal seed atoms.

Due to the overall neutrality of plasma and the conservation of electric change, the following equality applies:

$$n_e = n_i \tag{13}$$

Thus, Equation (12) can be written as

$$\frac{n_e^2}{n_s''} = C_{\text{Saha}} \, T^{1.5} \, e^{-\frac{\epsilon_i}{\theta}} \tag{14}$$

where the definition of ($C_{\text{Saha}}$) is provided in Equation (15).

$$C_{\text{Saha}} \equiv \left(\frac{2\pi \, m \, k}{h^2}\right)^{1.5} = 2.4146830 \times 10^{21} \left[\frac{1}{K^{1.5} \, m^3}\right] \tag{15}$$

Also, the sum of seed ions and seed atoms after ionization must be equal to the number of seed atoms before ionization (because the seed ions were initially seed atoms). Therefore, the following conservation relation in Equation (16) also applies:

$$n_s'' = n_s' - n_i \tag{16}$$

Replacing ($n_i$) by ($n_e$) in the above equation gives

$$n_s'' = n_s' - n_e \tag{17}$$

Using Equation (17), the number density of seed atoms after ionization ($n_s''$) in Equation (14) can be eliminated as shown in Equation (18).

$$\frac{n_e^2}{n_s' - n_e} = C_{\text{Saha}} \, T^{1.5} \, e^{-\frac{\epsilon_i}{\theta}} = K_{\text{Saha}} \tag{18}$$

or

$$\frac{n_e^2}{n_s' - n_e} = K_{\text{Saha}} \tag{19}$$

The above Equation (19) has a single unknown ($n_e$). The above equation can be arranged to take the standard form of a quadratic equation as shown in Equation (20).

$$a \, n_e^2 + b \, n_e + c = 0, \text{with } a = 1, b\left[\frac{1}{m^3}\right] = K_{\text{Saha}}, c\left[\frac{1}{m^6}\right] = K_{\text{Saha}} n_s' \tag{20}$$

The positive (non-trivial) root is the sought electron number density, thus

$$n_e\left[\frac{1}{m^3}\right] = \frac{-b + \sqrt{b^2 - 4\,a\,c}}{2a} = \frac{-K_{\text{Saha}} + \sqrt{K_{\text{Saha}}^2 - 4\,K_{\text{Saha}} n_s'}}{2} \tag{21}$$

It should be noted that the number density of electrons does not depend on the chemical composition of the carrier species. Thus, the same values are obtained regardless of the simulated fuel and oxidizer.

The number density of seed atoms before ionization ($n_s'$) needs to be computed first before the electron number density ($n_e$). This quantity ($n_s'$) can be calculated using the microscopic version of the ideal gas law as described in Equation (22) [127,128]

$$n_s'\left[\frac{1}{m^3}\right] = \frac{X_s' \, p_{tot}'}{k \, T} = \frac{p_s'}{k \, T} \tag{22}$$

In the present study, the initial total absolute pressure (before ionization but after seeding) in the simulations (unless otherwise specified) is 1 atm or 101325 Pa [129–131], and the mole reaction of the seed alkali metal is 0.01 or 1 % (also before ionization but after seeding). Thus, the seeding pressure ($p_s'$) is 1013.25 Pa by default.

### 2.3. Calculating the degree of ionization

The degree of ionization is defined as the percentage of seed atoms that were thermally ionized, which is also equal to the percentage of generated free electrons after thermal equilibrium ionization relative to the number of seed atoms that existed before ionization. In a mathematical form, it is defined as

$$\alpha[\%] \equiv \frac{n_i}{n_s'} \times 100\% = \frac{n_e}{n_s'} \times 100\% \tag{23}$$

It should be noted that the degree of ionization does not depend on the chemical composition of the carrier species. Thus, the same values are obtained regardless of the simulated fuel and oxidizer.

### 2.4. Calculating the mole fraction of ions and the total pressure after ionization

The number density of all neutrals before ionization (carriers and seed atoms) can be computed as

$$n_{tot}'\left[\frac{1}{m^3}\right] = \frac{p_{tot}'}{k \, T} = \frac{n_s'}{X_s'} \tag{24}$$

Due to the conservation of nuclei during the ionization process, this is also the number density of heavy particles (neutral carrier particles, seed atoms, and seed ions) after ionization. When adding the number density of electrons liberated during ionization, the result is the number density of all particles after ionization. Thus

$$n_{tot}''\left[\frac{1}{m^3}\right] = n_{tot}'\left[\frac{1}{m^3}\right] + n_e\left[\frac{1}{m^3}\right] \tag{25}$$

The mole fraction of electrons in the plasma after ionization is

$$X_e[-] \equiv \frac{n_e}{n_{tot}''} = \frac{n_{tot}'' - n_{tot}'}{n_{tot}''} = 1 - \frac{n_{tot}'}{n_{tot}''} \tag{26}$$

Due to the additional free electron particles that appear in the plasma after ionization, the total pressure after ionization is slightly larger than the total pressure before ionization, according to the relation below in Equation (27)

$$\frac{p_{tot}''}{p_{tot}'} = \frac{n_{tot}''}{n_{tot}'} = \frac{n_{tot}''}{n_{tot}'' - n_e} \tag{27}$$

Dividing the numerator and denominator of the right-hand side by ($n''_{tot}$) gives Equation (28), which is

$$\frac{p''_{tot}}{p'_{tot}} = \frac{n''_{tot}/n''_{tot}}{n''_{tot}/n''_{tot} - n_e/n''_{tot}} = \frac{1}{1 - X_e} \quad (28)$$

It should be noted that the increase in total pressure due to ionization does not depend on the chemical composition of the carrier species.

### 2.5. Calculating the mean frequency of electrons collision by neutrals scattering

The neutrals scattering is highly influenced by the electron-neutral collision cross-sections. For each neutral species (including seed atoms that remain as atoms and are not ionized), the product of the electron-neutral collision cross-sections and the electron speed is described analytically as a function of the electron energy expressed in eV. These formulas were presented in the Frost model based on a review of experimental data in the literature, and are considered valid for equilibrium plasma up to 5000 K. The expressions were given for 13 neutral gaseous species (counting K and Cs as two separate species although they have a common analytical expression). Argon (Ar) was not assigned an analytical expression related to the product of its electron-neutral collision cross-sections and the electron speed. Although argon is not among the combustion plasma gases used in the main simulations conducted here, it is included in the supplementary simulations performed to assess the predictive performance of the model applied here. Thus, this information about the electron-neutral collision cross-section is still needed. The electron-neutral collision cross-section for argon as utilized here was obtained by fitting an eighth-degree (nine-coefficients) polynomial to a curve reported in a separate study. That curve was obtained by comparing theoretical values and experimental results, and it described the correlation between argon electron-neutral collision cross-section and the electron energy in eV [132]. For consistency with the analytic expressions in the Frost model (where the product of the electron-neutral collision cross-section and the electron speed are treated as a single combined variable, not the electron-neutral collision cross-section by itself); for argon, the obtained polynomial fitting function is then multiplied by the mean (according to the Maxwell-Boltzmann speed distribution) electron speed, which depends only on the absolute temperature of the electrons (which is equal to the absolute temperature of the equilibrium plasma). This mean speed has the form

$$v\left[\frac{m}{s}\right] = C_v \sqrt{T} = \left(\frac{8\,k}{\pi\,m}\right)^{0.5} \sqrt{T} = 6212.511428620\,\sqrt{T[K]} \quad (29)$$

Because the published curve for the electron-neutral collision cross-section of argon versus electron energy in eV was in a log-log scale, the fitting was for the natural logarithm of the electron-neutral collision cross-section ($Q$ in Å$^2$) versus the natural logarithm of the electron energy ($u$ in eV). This logarithmic transformation of regression training data was useful in suppressing wiggles and improving the fitting accuracy compared to the case of attempting to model the raw (untransformed) data directly. The fitting was based on 23 data points taken along the curve with narrower spacing at the intermediate region of a steep variation but wider spacing at both ends, where the curve has a smoother variation. Fig. 2 shows the 23 transformed training data points (used to find the nine polynomial fitting coefficients) and the obtained fitting curve. Excellent agreement is evident. Lower-degree fitting polynomial functions (sixth-degree and seventh-degree were attempted also, but due to noticeable oscillation in their behavior, they were discarded and the eighth-degree fitting was selected after finding it free from such erratic oscillations. The fitting process was completed using the SciPy software library as an extension for the programming language Python to allow advanced scientific computing [133–138]. SciPy provides algorithms for optimization, and the "curve_fit" algorithm is the one that was used here. The SciPy version used is 1.6.2, and the Python version used is 3.8.8.

Fig. 3 shows the training data and the fitting curve when the logarithmic transformation is reversed, by taking the exponential of both ln (Q) and ln (u). The matching between the fitting curve and the discrete training data point is still apparent.

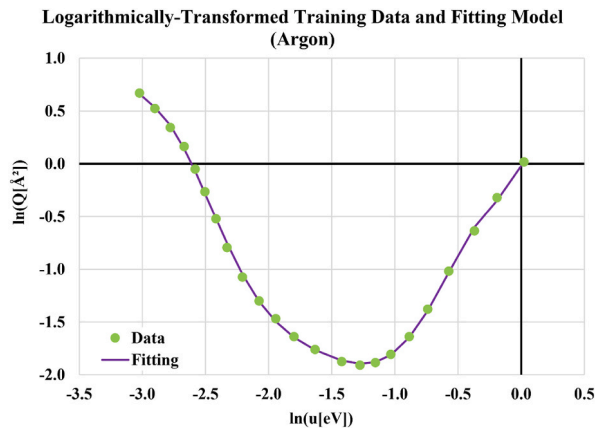

**Fig. 2.** Performance of the eighth-degree polynomial fitting of the natural logarithm of the electron-neutral collision cross-section of argon in Å$^2$, as a function of the natural logarithm of the electron energy in eV.

The analytic expressions for $Qv(u[eV])$ for 14 species (13 species provided in the Frost model, plus argon whose expression is proposed here), are listed in Table 2. The species are ordered alphabetically by the chemical symbol. Each analytical expression gives the product $(Qv)$ in m$^3$/s. For argon (Ar), the exponential term appearing in the analytical expression in the table gives the fitted electron-neutral collision cross-section in the very small unit of (Å$^2$), which explains the multiplicative factor of $10^{-20}$, which is added to adjust the units. The multiplied quantity (6212.511428620 $\sqrt{T[K]}$) is the mean electron speed in m/s (as a function of the absolute temperature), as given in Equation (29). For the remaining 13 species, the part of the expression between parentheses for each species gives $(Qv)$ in the very small unit of ($10^{-8}$ cm$^3$/s = $10^{-14}$ m$^3$/s), which explains the multiplicative factor of $10^{-14}$, which is added to adjust the units.

Using the proper set of analytical expressions (depending on the neutral species existing in the plasma gas), the mean collision cyclic frequency due to neutrals scattering is computed as the sum of the individual products of the number density of the neutral species (including non-ionized seed atoms) and the species $(Qv)$ quantity, as shown in Equation (30) below

$$\nu_0 \left[\frac{1}{s}\right] = \sum_{j}^{\text{neutrals}} n_j \left[\frac{1}{m^3}\right] (Qv)_j \left[\frac{m^3}{s}\right] \quad (30)$$

### 2.6. Calculating the mean frequency of electrons collision by coulomb scattering

The Coulomb scattering of electrons (their interaction with positively-charged or negatively-charged particles, being ions and other electrons) is quantified into a respective collision frequency $\nu_1$, through a series of steps. First, the absolute temperature of electrons (thus, the absolute temperature of the equilibrium plasma) is expressed as an energy term in eV [139–141], which is referred to here as the "thermal energy", described in Equation (31) below

$$\theta[eV] = \widehat{k}\, T \quad (31)$$

Then, an intermediate constant $(C_\Lambda)$ is computed as shown in Equation (32).

$$C_\Lambda \left[\frac{1}{V^{1.5}\, m^{1.5}}\right] \equiv \frac{3}{\sqrt{4\pi}\,(C_{\varepsilon 0}\, e)^{1.5}} = 1.5487761 \times 10^{13} \quad (32)$$

The above constant is used when computing the cutoff parameter $(\Lambda)$, according to Equation (33).

$$\Lambda = C_\Lambda \frac{\theta^{1.5}}{\sqrt{n_e}} \quad (33)$$

Another intermediate constant $(C_{K\nu 1})$ is computed as shown in Equation (34).

$$C_{K\nu 1} \left[\frac{V^2\, m^2\, C^{0.5}}{kg^{0.5}}\right] \equiv \sqrt{\frac{8}{m}}\, \pi\, e^{2.5}\, C_{\varepsilon 0}^2 = 7.72696841 \times 10^{-12} \quad (34)$$

This constant is combined with the effect of the temperature; the electron density; and the Coulomb logarithm, $\ln(\Lambda)$ [142–145]; to compute the mean or effective collision cyclic frequency due to Coulomb scattering as shown in Equation (35).

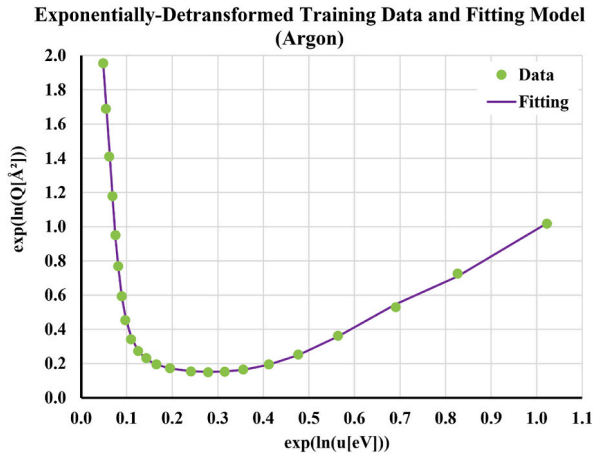

**Fig. 3.** Performance of the eighth-degree polynomial fitting when the fitting curve and the discrete data points are compared after counterbalancing the logarithmic transformation through an opposite exponential detransformation

**Table 2**
Analytical expressions for the product of the electron-neutral collision cross-section and the electron speed.

| Index | Gaseous Species | Analytic Expression for $Qv(u)$ in m$^3$/s ($u$ in eV) |
|---|---|---|
| 1 | Argon (Ar) | $10^{-20} \times 6212.511428620 \sqrt{T[K]} \times \exp\{-0.04488799052538687 + 2.6849940501234304 \ln(u) + 10.948947835474213 \ln(u)^2 + 34.755413381759915 \ln(u)^3 + 48.31120748606025 \ln(u)^4 + 34.44165112788354 \ln(u)^5 + 13.332151177313404 \ln(u)^6 + 2.662961571532757 \ln(u)^7 + 0.2148246953236797 \ln(u)^8\}$ |
| 2 | Carbon Monoxide (CO) | $10^{-14} \times (9.1\,u)$ |
| 3 | Carbon Dioxide (CO$_2$) | $10^{-14} \times \left(\frac{1.7}{\sqrt{u}} + 2.1\sqrt{u}\right)$ |
| 4 | Cesium Vapor (Cs) | $10^{-14} \times (160)$ |
| 5 | Atomic Hydrogen (H) | $10^{-14} \times (42\sqrt{u} - 14\,u)$ |
| 6 | Molecular Hydrogen (H$_2$) | $10^{-14} \times (4.5\sqrt{u} + 6.2\,u)$ |
| 7 | Water Vapor (H$_2$O) | $10^{-14} \times \left(\frac{10}{\sqrt{u}}\right)$ |
| 8 | Helium (He) | $10^{-14} \times (3.14\sqrt{u})$ |
| 9 | Potassium Vapor (K) | $10^{-14} \times (160)$ |
| 10 | Nitrogen (N$_2$) | $10^{-14} \times (12\,u)$ |
| 11 | Neon (Ne) | $10^{-14} \times (1.15\,u)$ |
| 12 | Atomic Oxygen (O) | $10^{-14} \times (5.5\sqrt{u})$ |
| 13 | Molecular Oxygen (O$_2$) | $10^{-14} \times (2.75\sqrt{u})$ |
| 14 | Hydroxyl Radical (OH) | $10^{-14} \times \left(\frac{8.1}{\sqrt{u}}\right)$ |

$$\nu_1\left[\frac{1}{s}\right] = \frac{0.952}{3} C_{K\nu 1} \frac{n_e \ln(\Lambda)}{\theta^{1.5}} \tag{35}$$

*2.7. Calculating the electron mobility and the electric conductivity*

After computing both frequencies of electron scattering modes of interaction (neutrals and Coulomb), the overall mean time between electron collisions is computed as given in Equation (36).

$$\tau[s] = \frac{1}{\nu} = \frac{1}{\nu_0 + \nu_1} \tag{36}$$

The electron mobility is the acquired drift speed (not orbital speed) by an electron when it is subject to a unit electric field, 1 V/m [146–148]. The electron mobility is computed as given in Equation (37).

$$\mu\left[\frac{m/s}{V/m}\right] = \frac{e}{m}\tau = \frac{e}{m\,\nu} \tag{37}$$

Finally, the estimated plasma electron conductivity is computed as given in Equation (38).

$$\sigma\left[\frac{S}{m}\right] = e\,n\,\mu = \frac{e^2 n}{m\,\nu} = \frac{e^2 n\,\tau}{m} \tag{38}$$

### 3. Assessing the electric conductivity model

This section focuses on evaluating the ability of the presented physical-mathematical model in the previous section to estimate the electric conductivity of partially-ionized plasma for MHD applications. To this end, a number of comparisons between model-derived results and published results are made, and the level of agreement is inspected. Before proceeding, it is important to mention that either modeling or measuring the plasma gas electric conductivity may involve large uncertainties [149–152]. Thus, the aim is not to reach a close agreement. Instead, the reasonableness of the modeling becomes accepted under qualitative broad consistency and correct profile of variations within the range of variables considered. In the assessment comparisons and in all main results to be discussed later, the ionization energy of cesium is set at $\epsilon_i$ (Cs) = 3.893 eV, and the ionization energy of cesium is set at $\epsilon_i$ (K) = 4.34 eV. These ae the values mentioned in the Frost model, which are proper [153–158].

*3.1. Comparison for electron density versus temperature*

Fig. 4 shows the predicted electron density for temperatures from 2000 K to 4000 K, which is the range for which similar results were reported with the Frost model. There are three seeding pressures for cesium in the figure, namely 0.001 atm (101.325 Pa), 0.02 atm (2026.5 Pa), and 0.04 atm (4053 Pa). The published results are superimposed in the figure for each seeding pressure, and the matching is excellent. This indicates correctly solving the Saha equilibrium ionization equation in the computations performed here. As mentioned earlier regarding Equation (21), the specification of the exact chemical composition of the carrier gases (the combustion product gases) is redundant in this stage of computation. To make this clearer, the definition of the thermal energy, $\theta[\text{eV}] = \widehat{k}\, T$, can be used in Equation (12) for $K_{\text{Saha}}$, which is then used in Equation (21) to manifest the dependence of the electron number density ($n_e$) only on: (1) the absolute temperature ($T[\text{K}]$), (2) the number density of seed atoms before ionization ($n'_s$), and (3) the ionization energy of the seed gas ($\epsilon_i[\text{eV}]$). This mathematical processing leads to a modified version of Equation (21), which is Equation (39).

$$n_e \left[\frac{1}{\text{m}^3}\right] = \frac{-C_{\text{Saha}}\, T^{1.5}\, e^{-\frac{\epsilon_i}{k\, T}} + \sqrt{C_{\text{Saha}}^2\, T^3\, e^{-\frac{2\,\epsilon_i}{k\, T}} + 4\, C_{\text{Saha}}\, T^{1.5}\, e^{-\frac{\epsilon_i}{k\, T}}\, n'_s}}{2} \qquad (39)$$

As can be seen in the above equation, the electron number density ($n_e$) is not impacted by the chemical composition of the carrier gases (whose ionization is neglected).

### 3.2. Comparison for electric conductivity versus temperature

Fig. 5 shows the predicted electrical conductivity over temperatures from 2000 K to 4000 K for three cesium-seeded noble gases. In these three cases, the total pressure is 1 atm (101325 Pa) before adding the seed, not after seeding as in the other simulations later. This exception is done to ensure matching with the simulation conditions reported by Frost, who published electric conductivity results for these three cases. The three noble gases are argon (with 101.325 Pa added Cs pressure), neon (with also 101.325 Pa added Cs pressure), and helium (with 2026.5 Pa added Cs pressure). Thus, the total absolute pressures after seeding in these three cases are 101426.325 Pa, 101426.325 Pa, and 103351.5 Pa; respectively. Despite the presence of deviations between the two sets of predictions, there is no systematic error such that the predictions here are neither always above nor always below the published profiles. Instead, at the lower temperatures part of the range, the implemented model here exceeds the published values, which is reversed near the higher temperatures part. Also, the implemented model here has a qualitative agreement with the published data in terms of the profile of variation with the temperature. For each case, 11 points from the published Frost profile were plotted (including the point of the lower temperature limit of 2000 K, and the point of the upper temperature limit of 4000 K). These 11 points were not merely taken at equal temperature intervals of 200 K. Instead, they were manually selected such that there are more points at the region of steeper change in the electric conductivity (from 2000 K top 3000 K) but fewer points when the change becomes smoother (from 3000 K TO 4000 K). Therefore, the selection of these points was based on the ability to visually reproduce the profile of the published curve in the original work of Frost for each of the three curves. When the relative magnitude of deviation between the two electric conductivity predictions at these 11 points is computed (this relative deviation is the magnitude of the difference between the two predictions divided by the value corresponding to the Frost prediction), the average was found to be 17.6 % for argon (Ar), 20.8 % for neon (Ne), and 13.2 % for helium (He). These numerical estimates of average deviation are considered mild.

### 3.3. Comparison for electron mole fraction and electric conductivity at a single condition

The third phase of comparisons corresponds to a potassium-seeded gas mixture representative of oxy-fuel combustion of methane ($O_2$–$CH_4$), with a relatively high absolute temperature of 3040 K. The details about this case and the independent results were reported in a study led by the National Energy Technology Laboratory (NETL) of the U.S. Department of Energy [159–162], which was concerned with modeling the electric conductivity of plasma and modeling the electron-ion collision cross-sections. Nine species with very insignificant fractions were omitted ($HO_2$, $H_2O_2$, $K^+$, $KH$, $KO$, $KOH$, $O^-$, and $OH^-$, and electrons) in the simulation here, where the individual mole fractions of these neutral or ionic species are approximately below $10^{-3}$. For the remaining more important nine species, the mole fractions and corresponding mass fractions were normalized to ensure that the sum is 1.0. The seed mass fraction of potassium was 1 % by mass, and this is incorporated here. The chemical composition of these species (the carrier gases and the seed potassium) is illustrated in Fig. 6. The total pressure was assumed 1 atm (before ionization but with the seeding potassium species included).

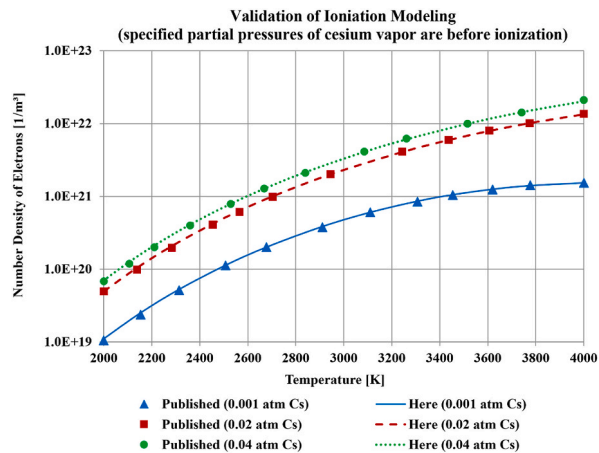

**Fig. 4.** Comparison of how the number density of electrons changes with the temperature for different seeding pressures of cesium, as published with the Frost model and as computed here.

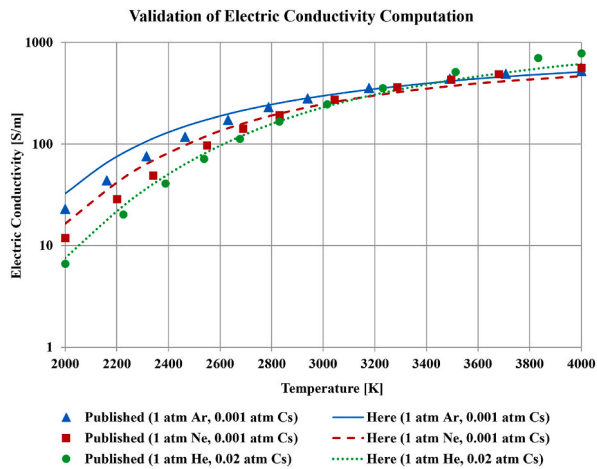

**Fig. 5.** Comparison of how the electric conductivity of seeded plasma changes with the temperature for different noble gases with specified seeding pressures of cesium, as published with the Frost model and as computed here.

Table 3 compares the predicted mole fraction of electrons in the independent study and the calculations here. The gap is viewed as tolerable.

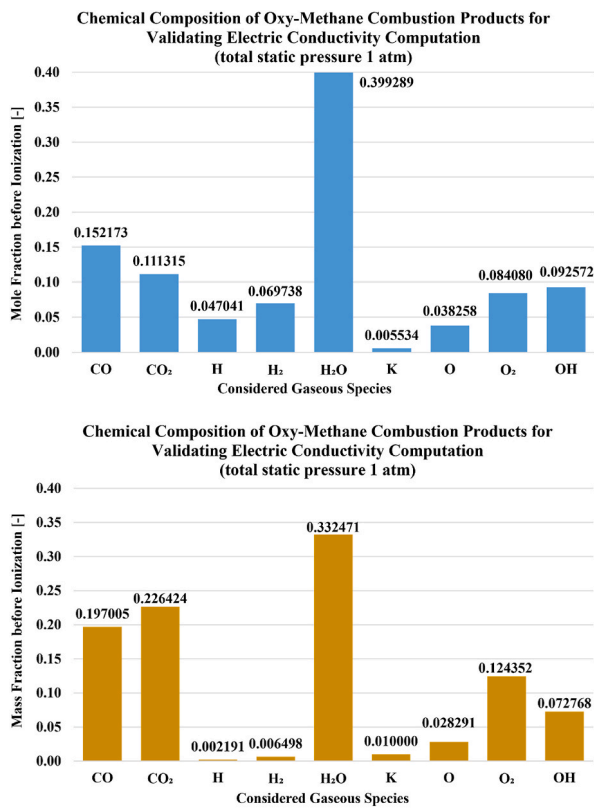

**Fig. 6.** Chemical composition used here for representing potassium-seeded (1 % mass fraction) oxygen-methane combustion products at 3040 K before ionization (top: mole fractions, bottom: mass fractions). The sum of listed mole fractions or mass fractions is adequately 1.000000.

Table 4 compares the electric conductivity obtained here (51.72 S/m) with the ones reported by NETL using electron-neutrals collision cross-section ($Q$) data recommended by NETL, and using other electron-neutrals collision cross-section data of three other studies. This table helps in demonstrating the large disparity possible when modeling the electric conductivity. The value reported here lies within the range of predictions, which makes it acceptable.

## 4. Results

In the current section, the impacts of the temperature, seed type, and chemical composition of the combustion products mixture are provided. In all the simulations to be presented, the total pressure before ionization (but after seeding) is 1 atm. The mole fraction of the seed vapor alkali metal before ionization ($X'_s$) is 0.01 (1 %). The seed elements considered in the analysis are cesium (Cs) and potassium (K).

### 4.1. Electron density

The first set of computation results presented here is about the electron density as a plasma property dependent on the temperature and seed type. This set is independent of the chemical composition of the plasma carrier gases.

Table 5 lists the values of the electron density in the considered temperature range from 2000 K to 3000 K (with a step of 100 K), while contrasting those values obtained when potassium (K) is the seed element and the values obtained when cesium (Cs) is the seed element. As expected, due to the lower ionization energy of cesium than potassium, the former corresponds to a higher electron density than the latter. The boosting in electron density in the case of cesium as the temperature increases from 2000 K to 3000 K is smaller than the one encountered with potassium. For cesium, this boost is about two-thirds of that of potassium. Thus, cesium is accompanied by less sensitivity of the electron density to the temperature than potassium.

The gain in electron density with cesium seeding instead of potassium seeding over the analyzed temperature range is illustrated in Fig. 7. This gain declines gradually as the temperature increases, from 3.66 at 2000 K to 2.33 at 3000 K. Due to the fixed number density of seed atoms (regardless of the seed type) before ionization in the current simulations, this gain is exactly equal to the gain in the degree of ionization ($\alpha$).

### 4.2. Degree of ionization and electrons mole fraction

The second set of computation results presented here is about the degree of ionization ($\alpha[\%]$) and the electrons mole fraction after the equilibrium thermal ionization of the seed atoms ($X_e$). Both quantities are dimensionless, making their interpretation more straightforward than the electron number density. Also, both quantities are independent of the specific chemical composition of the plasma carrier gases.

Fig. 8 contrasts the nonlinear growth of either the degree of ionization or the electrons mole fraction with the temperatures, for the case of cesium seeding and the case of potassium seeding. In the case of cesium, the degree of ionization at 3000 K is 69.26 times its value at 2000 K. In the case of potassium, the degree of ionization at 3000 K is 108.80 times its value at 2000 K. The corresponding boosting ratios for the electrons mole fractions are very similar, being 69.21 for cesium, and 108.77 for potassium. These big similarities in the dependence of the degree of ionization and the electrons mole fraction on the temperature are explained by the small presence of ions (thus, the small value of the electrons mole fractions). From Equations (23) and (26), a relation between the degree of ionization as a percentage ($\alpha[\%]$) and the electrons mole fraction ($X_e$) can be derived as given in Equation (40).

$$X_e = \frac{n_e}{n''_{tot}} = \frac{n_e}{n'_s} \frac{n'_s}{n''_{tot}} = \frac{n_e}{n'_s} \frac{n'_s}{n'_{tot}} \frac{n'_{tot}}{n''_{tot}} = \frac{n_e}{n'_s} X'_s (1 - X_e) = \frac{\alpha[\%]}{100} X'_s (1 - X_e) \tag{40}$$

Thus, the following relation in Equation (41) can be obtained:

$$\alpha[\%] = \frac{100 \, X_e}{X'_s (1 - X_e)} \tag{41}$$

The adopted seed mole fraction (before ionization) in the current investigative simulations is $X'_s = 0.01$. In addition, at small values of ($X_e$), as in the presented results, $(1 - X_e) \approx 1$. Therefore, the above equation can be simplified to the form in Equation (42).

$$\alpha[\%] \approx 10^4 \, X_e \tag{42}$$

This explains the found strong correlation between ($\alpha$) and ($X_e$). The magnitude of the relative error in the above ($\alpha, X_e$) approximate linear proportionality is $(1 - 1/(1 - X_e))$. Thus, the maximum error in the above linear proportionality between ($\alpha$) and ($X_e$) occurs at the largest value of ($X_e$), which corresponds to $T = 3000$ K with cesium seeding. This largest obtained $X_e(3000 \text{ K}, \text{Cs})$ is $6.6055 \times 10^{-4}$, making the worst linearity error less than $7 \times 10^{-4}$, which is negligible.

**Table 3**
Comparison of the electrons' mole fraction ($X_e$) for potassium-seeded (1 % mass fraction) oxygen-methane plasma at 3040 K.

| $X_e$ Published by NETL [–] | $X_e$ Computed Here [–] | $X_e$ Difference Magnitude [–] |
| --- | --- | --- |
| $1.84 \times 10^{-4}$ | $2.38 \times 10^{-4}$ | $5.4 \times 10^{-5}$ |

**Table 4**
Comparison of the plasma electric conductivity ($\sigma$) for potassium-seeded (1 % mass fraction) oxygen-methane plasma at 3040 K.

| Source Data of Q | $\sigma$ by NETL [S/m] | $\sigma$ Difference Magnitude \|Published – Here (51.72 S/m)\| [S/m] |
|---|---|---|
| NETL | 55.74 | 4.02 |
| Frost | 35.28 | 16.44 |
| Bünde et al. (1975) [163] | 22.13 | 29.59 |
| Spencer and Phelps (1976) [164] | 31.08 | 20.64 |
| Rosa (1987) [165] | 38.80 | 12.92 |

**Table 5**
Comparison of the variation of computed electron density ($n_e$) with the temperature, with two different seed materials.

| Temperature [K] | $n_e$ with $X'_s(K) = 0.01$ [1/m$^3$] | $n_e$ with $X'_s(Cs) = 0.01$ [1/m$^3$] |
|---|---|---|
| 2000 | $9.5783 \times 10^{18}$ | $3.5021 \times 10^{19}$ |
| 2100 | $1.7657 \times 10^{19}$ | $6.0678 \times 10^{19}$ |
| 2200 | $3.0803 \times 10^{19}$ | $1.0003 \times 10^{20}$ |
| 2300 | $5.1216 \times 10^{19}$ | $1.5791 \times 10^{20}$ |
| 2400 | $8.1643 \times 10^{19}$ | $2.3995 \times 10^{20}$ |
| 2500 | $1.2539 \times 10^{20}$ | $3.5248 \times 10^{20}$ |
| 2600 | $1.8633 \times 10^{20}$ | $5.0239 \times 10^{20}$ |
| 2700 | $2.6881 \times 10^{20}$ | $6.9685 \times 10^{20}$ |
| 2800 | $3.7762 \times 10^{20}$ | $9.4305 \times 10^{20}$ |
| 2900 | $5.1786 \times 10^{20}$ | $1.2478 \times 10^{21}$ |
| 3000 | $6.9475 \times 10^{20}$ | $1.6170 \times 10^{21}$ |
| Ratio (at 3000 K to at 2000 K) | 72.53 | 46.17 |

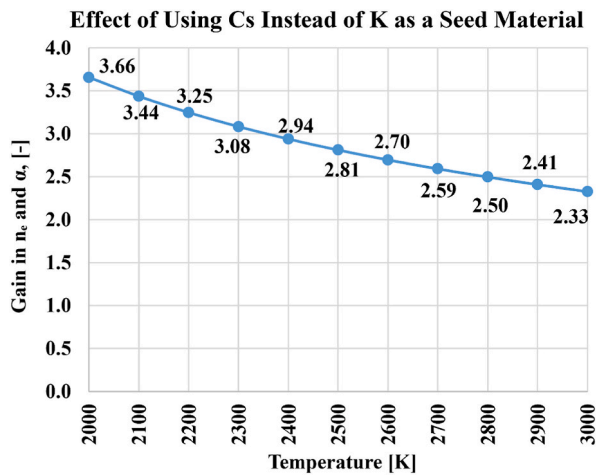

**Fig. 7.** Gain in the electron density or degree of ionization over the temperature range from 2000 K to 3000 K, if cesium (Cs) is used instead of potassium (K) as the seeded vapor.

This correlation between the degree of ionization (and similarly the electrons mole fraction) and the absolute temperature is close to an exponential function, as identified in Fig. 9 where a logarithmic scale is used for the degree of ionization and the electrons mole fraction, with a superimposed fitting curve. This is the same type of correlation, regardless of the seeding type.

### 4.3. Electric conductivity, oxy-fuel combustion plasma

The third set of computation results presented here is about the electric conductivity, for each of the two seed types, and for three different scenarios of combustion as reflected in the gaseous composition of the carrier gaseous species. Each chemical composition of the three scenarios mimics a stoichiometric complete combustion of a fuel with pure molecular oxygen as the oxidizer. These fuels are molecular hydrogen ($H_2$), methane ($CH_4$), and carbon (C). The first and third fuels represent the two extreme cases of carbon/hydrogen elemental mass ratio, being zero in the case of hydrogen ($H_2$), but infinity in the case of carbon (C). The second simulated fuel is the chief component of natural gas, forming more than 90 % of its content [166–169].

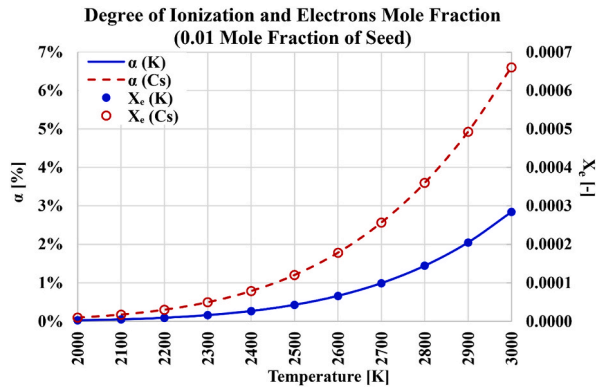

**Fig. 8.** Degree of ionization and electrons mole fraction over the temperature range from 2000 K to 3000 K, with either cesium (Cs) or potassium (K) used as the seeded vapor.

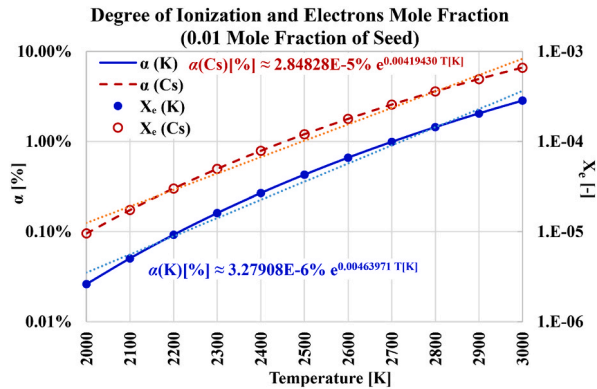

**Fig. 9.** Demonstration of the nearly exponential correlation between the degree of ionization and the electrons mole fraction with the absolute temperature, with either cesium (Cs) or potassium (K) used as the seeded vapor. The fitting curves have a dotted pattern. The corresponding repression equations are displayed.

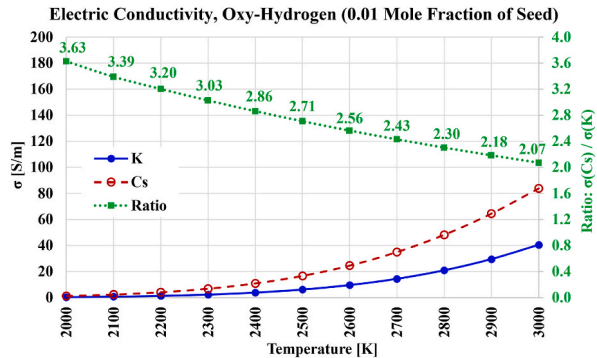

**Fig. 10.** Electric conductivity over the temperature range from 2000 K to 3000 K for the case of stoichiometric oxygen-hydrogen-based plasma composition, with either cesium (Cs) or potassium (K) used as a seed vapor. The ratio of both electric conductivities is also shown.

The combustion reactions for these selected three fuels are given in Equations 43–45 [170–172].

$$2\ H_2 + O_2 \rightarrow 2\ H_2O \tag{43}$$

$$CH_4 + 2\ O_2 \rightarrow CO_2 + 2\ H_2O \tag{44}$$

$$C + O_2 \rightarrow CO_2 \quad (45)$$

The chemical composition of the carrier gases (the combustion products) after the seed alkali metal is added can be described in terms of the mole fractions, which are given in Table 6.

The corresponding mass fractions of the neutral gas mixture before thermal equilibrium ionization of the seed component are listed in Table 7 for the case of potassium seeding, and in Table 8 for the case of cesium seeding.

The computed electric conductivities in the case of the oxy-hydrogen scenario are shown in Fig. 10, for the temperature range from 2000 K to 3000 K, and for each type of seed material. The gain in electric conductivity (due to using cesium instead of potassium as a seed material) is also shown. This gain drops gradually, in a manner close to linear, from 3.63 at 2000 K to 2.07 at 3000 K. In order to reach an electric conductivity of 20 S/m as an arbitrary target, a temperature of 2800 K is needed in the case of potassium seeding. However, a lower temperature of 2550 K is sufficient to reach this value. Below 2150 K (with potassium seeding) or below 1950 K (with cesium seeding), the electric conductivity does not even reach 1 S/m. This emphasizes the inefficacy of direct power extract through MHD channels if the temperature is not higher than typical values of fuel-fired boilers, where the flame temperature is generally below 1800 K [173]. Air preheating and oxygen-enriched oxidizers are two approaches to elevate the combustion temperatures [174,175].

The computed electric conductivities, as well as the electric conductivity gains due to changing the seed type from potassium to cesium, for the case of oxy-methane plasma, are shown in Fig. 11. The decline in the gain is similar to the one found for the oxy-hydrogen plasma. The reachable electric conductivities are improved in the case of oxy-methane compared to oxy-hydrogen, where the electric conductivity reaches 101.98 S/m at 3000 K with cesium seeding (compared to 83.78 S/m in oxy-hydrogen), and reaches 50.69 S/m at 3000 K with potassium seeding (compared to 40.51 S/m in oxy-hydrogen). With oxy-methane plasma and 1 % mole fraction of pre-ionization seeded cesium, a temperature of about 2500 K is sufficient to achieve an electric conductivity of 20 S/m, while a higher temperate of approximately 2750 K is needed if seeding is potassium.

The computed electric conductivities, as well as the electric conductivity gains due to changing the seed type, for the case of oxy-carbon plasma, are shown in Fig. 12. The decline in the gain is again close to a linear function in temperature, dropping from 3.59 at 3000 K to 1.77 at 2000 K. The reachable electric conductivities are remarkably improved in the case of oxy-carbon compared to oxy-hydrogen and oxy-methane, where the electric conductivity reaches 180.30 S/m at 3000 K with cesium seeding (compared to 83.78 S/m in oxy-hydrogen and 101.98 S/m in oxy-methane), and reaches 101.89 S/m at 3000 K with cesium seeding (compared to 40.51 S/m in oxy-hydrogen and 50.69 S/m in oxy-methane). With oxy-carbon plasma and 1 % mole fraction seeded cesium, a temperature of 2300 K is sufficient to achieve an electric conductivity of 20 S/m, while approximately 2500 K is needed if seeding is potassium.

### 4.4. Electric conductivity, air-fuel combustion plasma

The fourth set of computation results presented here is about the electric conductivity, for each of the two seed types, and for three additional scenarios of combustion that are the air-fuel version of the three combustion scenarios covered earlier. Air is modeled as a mixture of oxygen and nitrogen, with a molar ratio ($O_2:N_2$) of 1:3.762. Thus, the corresponding mole fractions are 0.2100 (21 %) for $O_2$ and 0.7900 (79 %) for $N_2$ [176–181]. The reactions in Equations 46–48 describe the stoichiometric complete combustion of molecular hydrogen, methane, and carbon; with air being the oxidizer. Compared with the oxy-fuel scenarios, the air-fuel combustion scenarios have an additional nitrogen species, which influences the electron-neutral collision cross-sections, and thus influences the electric conductivity.

$$2\,H_2 + O_2 + 3.762\,N_2 \rightarrow 2\,H_2O + 3.762\,N_2 \quad (46)$$

$$CH_4 + 2\,O_2 + 7.524\,N_2 \rightarrow CO_2 + 2\,H_2O + 7.524\,N_2 \quad (47)$$

$$C + O_2 + 3.762\,N_2 \rightarrow CO_2 + 3.762\,N_2 \quad (48)$$

Based on previously mentioned analytical expressions for the product of the electron-neutral collision cross-section and the electron speed, ($Qv$), its value for molecular nitrogen ($N_2$) is compared with its counterparts for water vapor ($H_2O$) and carbon dioxide ($CO_2$) in Table 9. The comparison is made at the temperatures 2000 K, 2500 K, and 3000 K. These three temperatures correspond to the beginning, middle, and end, of the temperature range of interest; respectively. For all temperatures, it is noticed that the electron-neutral collision cross-sections for nitrogen are much less than the electron-neutral collision cross-sections for water vapor. Consequently, electric conductivity should be improved when water vapor is replaced by nitrogen (while keeping all other properties the same). Compared with carbon dioxide, the electron-neutral collision cross-sections for nitrogen are of comparable magnitude. The electron-neutral collision cross-section for nitrogen is smaller between 2000 K and 2662 K, but this is reversed between 2663 K and 3000 K. Thus, the presence of nitrogen instead of carbon dioxide is not expected to cause substantial changes in the electrical

**Table 6**
Chemical compositions (expressed as mole fractions) of the pre-ionization seeded gas representing oxy-fuel combustion for three fuels.

| Fuel | $H_2O$ | $CO_2$ | K or Cs | Sum |
| --- | --- | --- | --- | --- |
| Hydrogen | 0.990000 | 0 | 0.010000 | 1.000000 |
| Methane | 0.660000 | 0.330000 | 0.010000 | 1.000000 |
| Carbon | 0 | 0.990000 | 0.010000 | 1.000000 |

**Table 7**
Chemical compositions (expressed as mass fractions) of the pre-ionization seeded gas representing oxy-fuel combustion for three fuels, in the case of potassium seeding.

| Fuel | $H_2O$ | $CO_2$ | K | Sum |
| --- | --- | --- | --- | --- |
| Hydrogen | 0.978548 | 0 | 0.021452 | 1.000000 |
| Methane | 0.44359059 | 0.54182279 | 0.01458662 | 1.00000000 |
| Carbon | 0 | 0.991106 | 0.008894 | 1.000000 |

**Table 8**
Chemical compositions (expressed as mass fractions) of the pre-ionization seeded gas representing oxy-fuel combustion for three fuels, in the case of cesium seeding.

| Fuel | $H_2O$ | $CO_2$ | Cs | Sum |
| --- | --- | --- | --- | --- |
| Hydrogen | 0.930649 | 0 | 0.069351 | 1.000000 |
| Methane | 0.428591 | 0.523502 | 0.047907 | 1.000000 |
| Carbon | 0 | 0.970399 | 0.029601 | 1.000000 |

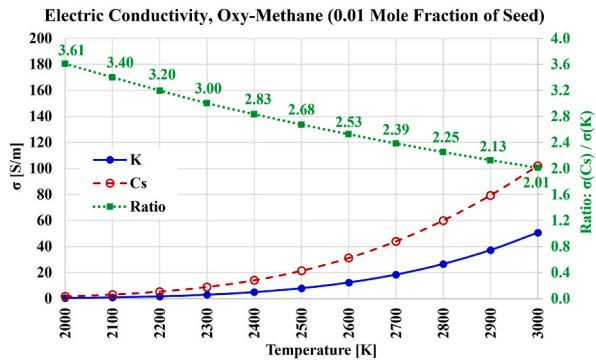

**Fig. 11.** Electric conductivity over the temperature range from 2000 K to 3000 K for the case of stoichiometric oxygen-methane-based plasma composition, with either cesium (Cs) or potassium (K) used as a seed vapor. The ratio of both electric conductivities is also shown.

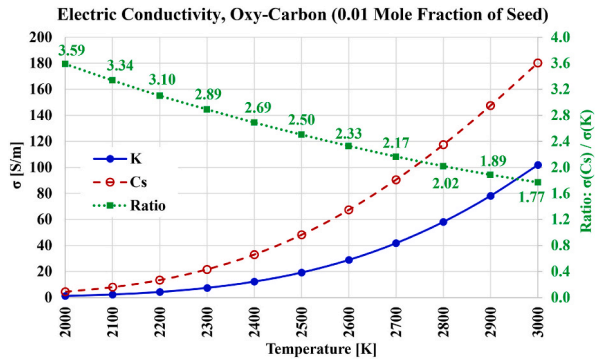

**Fig. 12.** Electric conductivity over the temperature range from 2000 K to 3000 K for the case of stoichiometric oxygen-carbon-based plasma composition, with either cesium (Cs) or potassium (K) used as a seed vapor. The ratio of both electric conductivities is also shown.

**Table 9**
The product of the electron-neutral collision cross-section and the electron speed, ($Qv$), for nitrogen in comparison with water vapor and with carbon dioxide.

| T [K] | $u$[eV] | $Qv\left[\frac{m^3}{s}\right]$, $H_2O$ | $Qv\left[\frac{m^3}{s}\right]$, $CO_2$ | $Qv\left[\frac{m^3}{s}\right]$, $N_2$ | $\frac{Qv(N_2)}{Qv(H_2O)}$ | $\frac{Qv(N_2)}{Qv(CO_2)}$ |
| --- | --- | --- | --- | --- | --- | --- |
| 2000 | 0.25852 | 1.9668E-13 | 4.4112E-14 | 3.1022E-14 | 0.158 | 0.703 |
| 2500 | 0.32315 | 1.7591E-13 | 4.1843E-14 | 3.8778E-14 | 0.220 | 0.927 |
| 3000 | 0.38778 | 1.6059E-13 | 4.0377E-14 | 4.6534E-14 | 0.290 | 1.152 |

conductivity, with small improvement near 2000 K but a small drop near 3000 K.

The chemical composition of the carrier gases mixed with the added seed vapor is explained in Table 10 (as mole fractions), in Table 11 (as mass fractions with potassium seed), and in Table 12 (as mass fractions with cesium seed); for each of the three air-fuel combustion scenarios.

The electric conductivity results for air-hydrogen combustion are shown in Fig. 13. The gain in electric conductivity due to the seed type has a similar profile to what was found in the case of oxy-hydrogen combustion, but with a slight reduction in this gain in the case of oxy-hydrogen over the entire range of temperatures from 2000 K to 3000 K. The electric conductivity increased with air-hydrogen combustion compared to oxy-hydrogen combustion. For example, at 3000 K; the electric conductivity with cesium seeding is 125.35 S/m with air-hydrogen combustion (compared to 83.78 S/m with oxy-hydrogen combustion); and the electric conductivity with potassium seeding is 64.63 S/m with air-hydrogen combustion (compared to 40.51 S/m with oxy-hydrogen combustion). For air-hydrogen combustion and cesium seeding, an electric conductivity of 20 S/m is attainable at about 2400 K (compared to about 2550 K with oxy-hydrogen combustion). For air-hydrogen combustion and potassium seeding, an electric conductivity of 20 S/m is attainable at about 2650 K (compared to about 2800 K with oxy-hydrogen combustion).

The electric conductivity results for air-methane combustion are shown in Fig. 14. The profile of the gains in electric conductivity due to the seed remained similar to the profile in the case of oxy-hydrogen combustion. Except close to 2000 K, this gain mildly drops in the case of oxy-methane combustion; with the gain becomes 1.89 at 3000 K in the case of air-methane combustion (compared to 2.01 at 3000 K in the case of oxy-methane combustion). The maximum reachable electric conductivity at 3000 K with cesium seeding is 142.97 S/m in the case of air-methane combustion (compared to 101.98 S/m in the case of oxy-methane combustion); and the electric conductivity with potassium seeding is 75.84 S/m in the case of air-methane combustion (compared to 50.69 S/m in the case of oxy-methane combustion). For air-methane combustion and cesium seeding, an electric conductivity of 20 S/m is attainable at about 2350 K (compared to about 2500 K in the case of oxy-methane combustion). For air-methane combustion and potassium seeding, an electric conductivity of 20 S/m is attainable at about 2600 K (compared to about 2750 K in the case of oxy-methane combustion).

The electric conductivity results for air-carbon combustion are shown in Fig. 15. The variation of the gain in electric conductivity due to using cesium instead of potassium is nearly the same as it was for oxy-carbon combustion, with the gain dropping nearly linearly as the temperature increases. This gain is 3.58 at 2000 K and 1.79 at 3000 K. The maximum reachable electric conductivity at 3000 K with cesium seeding is 172.27 S/m, and 96.00 S/m with potassium seeding. These are similar (but slightly lower) than their counterparts in the case of oxy-carbon combustion. Due to the similarity in the electric conductivity profiles in air-carbon combustion and oxy-carbon combustion, an electric conductivity of 20 S/m is attainable in the case of air-carbon combustion at about 2300 K with cesium seeding, and at about 2500 K with potassium seeding; which are the same approximate temperature values mentioned earlier in the case of oxy-carbon combustion. Thus, it can be said that when carbon dioxide is replaced by molecular nitrogen, no large changes occur in the electric conductivity.

To explain the reason for the mentioned particular feature of carbon dioxide (its weak sensitivity to the oxidizer type), its analytical expression for the product of the electron-neutral collision cross-section and the electron speed, $Qv(u[\text{eV}])$ in Tables 2 and is transformed using Equation (29), $v[\text{m/s}] = 6212.511428620\ \sqrt{T[\text{K}]}$, the used assumption for the temperature-dependent electron kinetic energy value $u[\text{eV}] = 1.5\ \widehat{k}_B[\text{eV/K}]\ T[\text{K}] = 0.00012925999893\ T[\text{K}]$, and the units conversion rule: $\text{m}^2 = 10^{20} \text{Å}^2$. This transformation leads to a direct relation between the mean electron-neutral collision cross-section $Q[\text{Å}^2]$ and the temperature $T[\text{K}]$, which is given in Table 13 for carbon dioxide and molecular nitrogen. This relation is visualized for either carrier gas in Fig. 16. Although the profile curve (variation of $Q[\text{Å}^2]$ with $T[\text{K}]$) for either gas resembles a straight line over the range of temperatures covered here (from 2000 K to 3000 K), it is actually slightly nonlinear. From the figure, it can be seen that the $Q$ values for these two gases (over the covered temperature range) have oppositve trends. While the $Q$ values decrease with temperature for carbon dioxide, they increase for nitrogen. Furthermore, the $Q$ values for nitrogen and carbon dioxide are not very different. They become equal at an intermediate temperature near 2650 K. Thus, replacing carbon dioxide (in oxy-carbon combustion) with nitrogen (in air-carbon combustion) does not cause significant changes in the resultant electric conductivity.

## 5. Discussion

After the analysis of electric conductivities was presented in the previous section, the current section aims to provide auxiliary details about the computed electric conductivities of plasma, for different chemical compositions of carrier gases, different alkali metal seed types, and different temperatures. One advantage of numerical modeling compared to measurements is the ability to find details that are not accessible experimentally. This advantage is employed here by reporting the two components of the electric conductivities that when combined (through harmonic averaging and then halving), the total electric conductivity is obtained. Although practically, the individual components are not important to know because their combined final value is what impacts the performance of direct power extraction in MHD channels; it still can be valuable to identify the relative size of both components at different plasma conditions.

**Table 10**
Chemical compositions (expressed as mole fractions) of the pre-ionization seeded gas mixture representing air-fuel combustion for three fuels.

| Fuel | $H_2O$ | $CO_2$ | $N_2$ | K or Cs | Sum |
| --- | --- | --- | --- | --- | --- |
| Hydrogen | 0.343631 | 0 | 0.646369 | 0.010000 | 1.000000 |
| Methane | 0.188141 | 0.094071 | 0.707788 | 0.010000 | 1.000000 |
| Carbon | 0 | 0.207896 | 0.782104 | 0.010000 | 1.000000 |

**Table 11**
Chemical compositions (expressed as mass fractions) of the pre-ionization seeded gas mixture representing air-fuel combustion for three fuels, in the case of potassium seeding.

| Fuel | $H_2O$ | $CO_2$ | $N_2$ | K | Sum |
|---|---|---|---|---|---|
| Hydrogen | 0.25074776 | 0 | 0.73341566 | 0.01583658 | 1.00000000 |
| Methane | 0.122150 | 0.149200 | 0.714559 | 0.014091 | 1.000000 |
| Carbon | 0 | 0.290921 | 0.696647 | 0.012432 | 1.000000 |

**Table 12**
Chemical compositions (expressed as mass fractions) of the pre-ionization seeded gas mixture representing air-fuel combustion for three fuels, in the case of cesium seeding.

| Fuel | $H_2O$ | $CO_2$ | $N_2$ | Cs | Sum |
|---|---|---|---|---|---|
| Hydrogen | 0.241569 | 0 | 0.706569 | 0.051862 | 1.000000 |
| Methane | 0.118156 | 0.144321 | 0.691192 | 0.046331 | 1.000000 |
| Carbon | 0 | 0.2824946 | 0.6764698 | 0.0410356 | 1.0000000 |

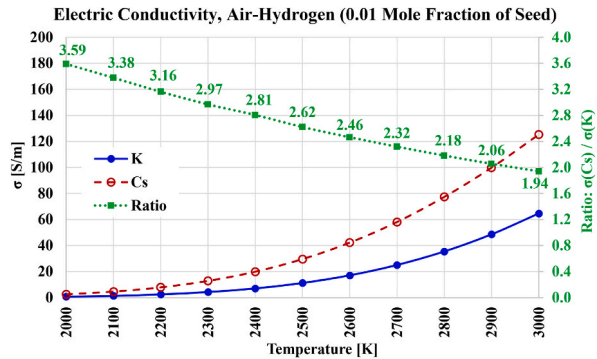

**Fig. 13.** Electric conductivity over the temperature range from 2000 K to 3000 K for the case of stoichiometric air-hydrogen-based plasma composition, with either cesium (Cs) or potassium (K) used as a seed vapor. The ratio of both electric conductivities is also shown.

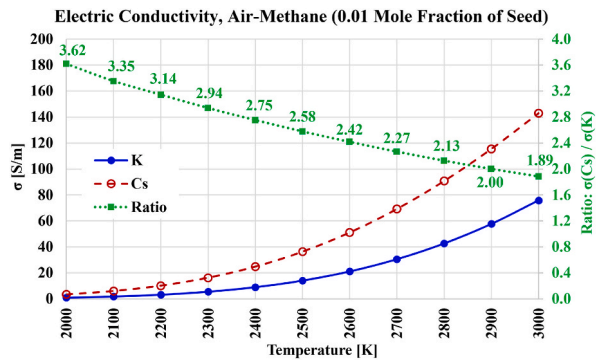

**Fig. 14.** Electric conductivity over the temperature range from 2000 K to 3000 K for the case of stoichiometric air-methane-based plasma composition, with either cesium (Cs) or potassium (K) used as a seed vapor. The ratio of both electric conductivities is also shown.

Table 14 lists the two components of electric conductivities (the neutrals-only components $\sigma_0$, and the Coulomb scattering component $\sigma_1$), the combined final electric conductivity ($\sigma$), the percentage of this combined final electric conductivity as a fraction of the neutrals-only component ($100\% \times \sigma/\sigma_0$), and the corresponding electrons mole fraction ($X_e$). This set of five quantities is given 36 times, which correspond to six chemical compositions (three oxy-fuel combustion scenarios and three other air-fuel combustion scenarios), two seed types (potassium, K; and cesium, Cs), and three temperatures ($T$ = 2000 K, 2500 K, and 3000 K).

Because the limiting factor in ($\sigma$) is the minimum of ($\sigma_0$) and ($\sigma_1$), the inspection of the enumerated values can help in identifying the threshold of electrons mole fraction at which the very-weakly-ionized plasma (where $\sigma_1$ is totally neglected) becomes suitable.

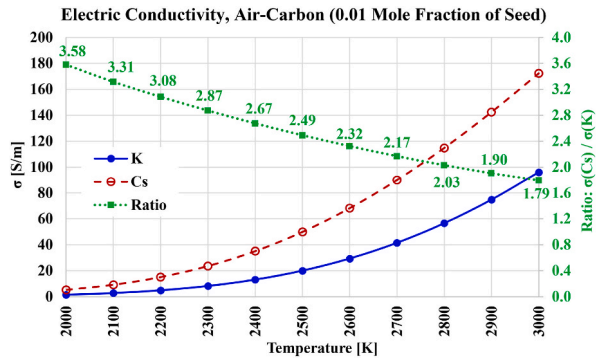

**Fig. 15.** Electric conductivity over the temperature range from 2000 K to 3000 K for the case of stoichiometric air-carbon-based plasma composition, with either cesium (Cs) or potassium (K) used as a seed vapor. The ratio of both electric conductivities is also shown.

**Table 13**
Transformed expression for the mean electron-neutral collision cross-section of carbon dioxide and nitrogen.

| Gaseous Species | Untransformed expression for $Qv(u)$ in m$^3$/s ($u$ in eV), from Table 2 | Equivalent (transformed) expression for $Q(T)$ in Å$^2$ ($T$ in K) |
| --- | --- | --- |
| Carbon Dioxide ($CO_2$) | $10^{-14} \times \left(\dfrac{1.7}{\sqrt{u}} + 2.1\sqrt{u}\right)$ | $\dfrac{24068.534}{T} + 3.8431220$ |
| Nitrogen ($N_2$) | $10^{-14} \times (12\,u)$ | $0.24967680\,\sqrt{T}$ |

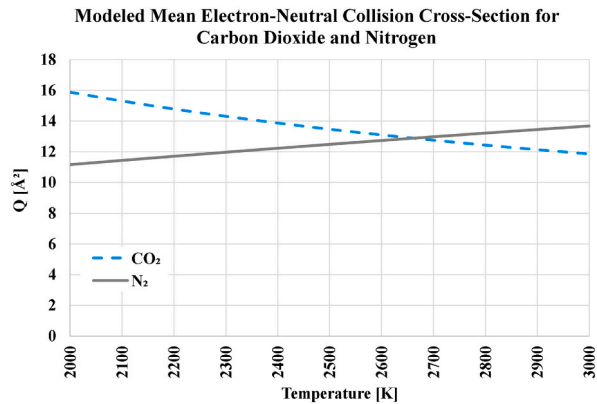

**Fig. 16.** Mean electron-neutral collision cross-section for carbon dioxide ($CO_2$) and molecular nitrogen ($N_2$) over the temperature range from 2000 K to 3000 K, according to the analytical expressions used here.

Because the values of the electrons mole fraction ($X_e$) are independent of the exact composition of the carrier gases, each ($X_e$) is repeated six times (same value for each of the six combustion scenarios). The values of ($X_e$) range from $2.6103 \times 10^{-6}$ (at 2000 K with potassium seeding) to $6.6055 \times 10^{-4}$ (at 3000 K with cesium seeding); thus, the values vary by two orders of magnitude (a maximum-to-minimum ratio of 253). The largest listed value of (100% × $\sigma/\sigma_0$) is 99.75 % (oxy-hydrogen combustion, potassium seeding, and 2000 K). This is where ($\sigma_1$) is most-negligible compared to ($\sigma_0$). The smallest listed value of (100% × $\sigma/\sigma_0$) is 53.16 % (oxy-carbon combustion, cesium seeding, and 3000 K). This is where ($\sigma_1$) is nearly equally important as ($\sigma_0$).

The results suggest that relying on ($X_e$) alone as a threshold for classifying the plasma as very-weakly-ionized or not is not accurate, because the type of seed and carrier gases is also important to take into consideration. For example, with a small ($X_e$) of $1.2006 \times 10^{-4}$ (2500 K, cesium seeding), the percentage (100% × $\sigma/\sigma_0$) can be relatively far from 100 % (80.84 % in the case of air-carbon combustion); while a larger ($X_e$) of $2.8392 \times 10^{-4}$ (3000 K, potassium seeding); that percentage can become closer to 100 % (88.34 % in the case of oxy-hydrogen combustion), which is opposite to the generic expectation for a decrease in this percentage as the electrons mole fraction increases.

Overall, a low threshold of about ($X_e$ [−] ≤ $10^{-5}$) appears to be a proper conservative criterion to neglect Coulomb scattering without a detrimental effect on the plasma electric conductivity.

Table 14

Differentiation between the two components of electric conductivity in relation to the electrons mole fraction; at 2000 K, 2500 K, and 3000 K; per seed type and combustion type.

| Combustion Type | Seed | T [K] | $X_e$ [−] | $\sigma_0 \left[\dfrac{S}{m}\right]$ | $\sigma_1 \left[\dfrac{S}{m}\right]$ | $\sigma \left[\dfrac{S}{m}\right]$ | $\dfrac{\sigma}{\sigma_0} \times 100\%$ |
|---|---|---|---|---|---|---|---|
| Oxy-Hydrogen | K | 2000 | $2.6103 \times 10^{-6}$ | 0.349 | 139.825 | 0.348 | 99.75 % |
| | | 2500 | $4.2713 \times 10^{-5}$ | 6.332 | 233.120 | 6.165 | 97.36 % |
| | | 3000 | $2.8392 \times 10^{-4}$ | 45.856 | 347.515 | 40.510 | 88.34 % |
| | Cs | 2000 | $9.5439 \times 10^{-6}$ | 1.276 | 157.147 | 1.266 | 99.19 % |
| | | 2500 | $1.2006 \times 10^{-4}$ | 17.812 | 260.422 | 16.672 | 93.60 % |
| | | 3000 | $6.6055 \times 10^{-4}$ | 107.094 | 384.916 | 83.783 | 78.23 % |
| Oxy-Methane | K | 2000 | $2.6103 \times 10^{-6}$ | 0.459 | 139.825 | 0.457 | 99.67 % |
| | | 2500 | $4.2713 \times 10^{-5}$ | 8.253 | 233.120 | 7.971 | 96.58 % |
| | | 3000 | $2.8392 \times 10^{-4}$ | 59.344 | 347.515 | 50.688 | 85.41 % |
| | Cs | 2000 | $9.5439 \times 10^{-6}$ | 1.677 | 157.147 | 1.660 | 98.94 % |
| | | 2500 | $1.2006 \times 10^{-4}$ | 23.220 | 260.422 | 21.319 | 91.81 % |
| | | 3000 | $6.6055 \times 10^{-4}$ | 138.738 | 384.916 | 101.981 | 73.51 % |
| Oxy-Carbon | K | 2000 | $2.6103 \times 10^{-6}$ | 1.233 | 139.825 | 1.222 | 99.13 % |
| | | 2500 | $4.2713 \times 10^{-5}$ | 20.986 | 233.120 | 19.253 | 91.74 % |
| | | 3000 | $2.8392 \times 10^{-4}$ | 144.149 | 347.515 | 101.887 | 70.68 % |
| | Cs | 2000 | $9.5439 \times 10^{-6}$ | 4.508 | 157.147 | 4.383 | 97.21 % |
| | | 2500 | $1.2006 \times 10^{-4}$ | 59.120 | 260.422 | 48.182 | 81.50 % |
| | | 3000 | $6.6055 \times 10^{-4}$ | 339.178 | 384.916 | 180.301 | 53.16 % |
| Air-Hydrogen | K | 2000 | $2.6103 \times 10^{-6}$ | 0.710 | 139.825 | 0.706 | 99.49 % |
| | | 2500 | $4.2713 \times 10^{-5}$ | 11.865 | 233.120 | 11.291 | 95.16 % |
| | | 3000 | $2.8392 \times 10^{-4}$ | 79.390 | 347.515 | 64.626 | 81.40 % |
| | Cs | 2000 | $9.5439 \times 10^{-6}$ | 2.595 | 157.147 | 2.553 | 98.38 % |
| | | 2500 | $1.2006 \times 10^{-4}$ | 33.394 | 260.422 | 29.599 | 88.63 % |
| | | 3000 | $6.6055 \times 10^{-4}$ | 185.884 | 384.916 | 125.350 | 67.43 % |
| Air-Methane | K | 2000 | $2.6103 \times 10^{-6}$ | 0.930 | 139.825 | 0.924 | 99.34 % |
| | | 2500 | $4.2713 \times 10^{-5}$ | 14.969 | 233.120 | 14.066 | 93.97 % |
| | | 3000 | $2.8392 \times 10^{-4}$ | 97.014 | 347.515 | 75.842 | 78.18 % |
| | Cs | 2000 | $9.5439 \times 10^{-6}$ | 3.400 | 157.147 | 3.328 | 97.88 % |
| | | 2500 | $1.2006 \times 10^{-4}$ | 42.143 | 260.422 | 36.273 | 86.07 % |
| | | 3000 | $6.6055 \times 10^{-4}$ | 227.454 | 384.916 | 142.970 | 62.86 % |
| Air-Carbon | K | 2000 | $2.6103 \times 10^{-6}$ | 1.488 | 139.825 | 1.472 | 98.95 % |
| | | 2500 | $4.2713 \times 10^{-5}$ | 21.902 | 233.120 | 20.021 | 91.41 % |
| | | 3000 | $2.8392 \times 10^{-4}$ | 132.645 | 347.515 | 96.001 | 72.37 % |
| | Cs | 2000 | $9.5439 \times 10^{-6}$ | 5.442 | 157.147 | 5.260 | 96.65 % |
| | | 2500 | $1.2006 \times 10^{-4}$ | 61.704 | 260.422 | 49.885 | 80.84 % |
| | | 3000 | $6.6055 \times 10^{-4}$ | 311.834 | 384.916 | 172.271 | 55.24 % |

## 6. Conclusions

The current study covered numerical modeling of the electric conductivity of seeded partially-ionized combustion-based equilibrium plasma for use in direct power extraction through magnetohydrodynamic (MHD) channels. The model was utilized to explore the electric properties of plasma as dependent on the seed type, the temperature, and the combustion type (fuel and oxidizer pair). The range of temperatures considered was from 2000 K to 3000 K, which is relevant to both air-fuel combustion and oxy-fuel combustion. The mole fraction of the seeded alkali metal vapor before ionization was 0.01 (1 %). The fuels considered in the simulations are molecular hydrogen, methane, and carbon; and the oxidizers considered are pure molecular oxygen and air (oxygen-nitrogen mixture). The plasma was considered to be atmospheric (having an absolute pressure of 1 atm) after adding the seed vapor (thus, 0.99 atm before the seeding).

The findings of this study suggest that.

- The electric conductivity is higher when the carrier gases have less content of water vapor or more content of carbon dioxide.
- Regardless of the alkali metal seed type (potassium or cesium) and the combustion type, an electric conductivity of 40 S/m can be exceeded if the temperature can reach 3000 K, and an electric conductivity of 20 S/m can be exceeded if the temperature can reach 2800 K.
- Carbon dioxide and molecular nitrogen have similar effects on the plasma electric conductivity.
- Cesium seeding can raise the electric conductivity to more than two or three times (depending on the temperature) its value when potassium seeding is used.
- Coulomb scattering may be neglected if the electrons mole fraction in the partially-ionized plasma is $10^{-5}$ (0.001 %) or less.

The current study can be extended in a variety of ways. One particular extension is the quantification of the effect of the carbon-to-hydrogen (C/H) ratio in the fuel (thus, the molar ratio of the carbon dioxide to water vapor in the pre-ionization combustion gases). In the present study, three values were covered, making this factor interpreted as a discrete variable or a parameter. If more values are added, and if a transformation is applied (to make this factor bounded rather than semi-infinite, proper identification of the nonlinear influence of this factor may be successfully attained.

Not applicable (this research has a single author).

## Funding

Not applicable (this research received no funding).

## Institutional review board statement

Not applicable (this research does not involve humans, animals, or environmental hazards).

## Informed consent statement

Not applicable (this research does not involve humans).

## Data availability statement

No data associated with the study have been deposited into a publicly available repository. Instead, the data are included in the article itself (please see the figures and tables in the Results section).

## CRediT authorship contribution statement

**Osama A. Marzouk:** Writing – review & editing, Writing – original draft, Visualization, Validation, Software, Methodology, Investigation, Formal analysis, Conceptualization.

## Declaration of competing interest

The authors declare that they have no known competing financial interests or personal relationships that could have appeared to influence the work reported in this paper.